\documentclass{article}
\usepackage{spconf,amsmath,graphicx,booktabs,pifont}
\usepackage{makecell}

\title{A robust audio deepfake detection system via multi-view feature}
\name{Yujie Yang\textdagger \thanks{\textdagger Euqal contribution}, Haochen Qin\textdagger\footnotemark[1], Hang Zhou, Chengcheng Wang, Tianyu Guo, Kai Han*\thanks{*Corresponding author}, Yunhe Wang*\footnotemark[2]}
\address{Huawei Noah's Ark Lab}
\begin{document}
%
\maketitle
\begin{abstract}
With the advancement of generative modeling techniques, synthetic human speech becomes increasingly indistinguishable from real, and tricky challenges are elicited for the audio deepfake detection (ADD) system. In this paper, we exploit audio features to improve the generalizability of ADD systems. Investigation of the ADD task performance is conducted over a broad range of audio features, including various handcrafted features and learning-based features. Experiments show that learning-based audio features pretrained on a large amount of data generalize better than hand-crafted features on out-of-domain scenarios. Subsequently, we further improve the generalizability of the ADD system using proposed multi-feature approaches to incorporate complimentary information from features of different views. The model trained on ASV2019 data achieves an equal error rate of 24.27\% on the In-the-Wild dataset.
The code will be released as soon $\footnote[1]{Mindspore: https://gitee.com/mindspore/models}
$.
\end{abstract}
\section{Introduction}
\label{sec:intro}

AI technology currently has made breakthroughs with the support of large-scale models, massive datasets, and powerful computing capabilities. 
Speech synthesis, speech conversion, and speech editing technologies have been able to generate human speech that is virtually indistinguishable from real human speech. 
However, progress in these speech generation technologies has also raised potential threats. 
The synthetic speech could be misused for spreading rumors, executing fraud, and eliciting other illicit activities.
Therefore, identifying synthetic speech is increasingly important. 
In response to such challenges, efforts including automatic speaker verification spoofing and countermeasures (ASVspoof), and audio deepfake detection (ADD) competitions
have been held to collect solutions \cite{todisco2019asvspoof,yamagishi2021asvspoof,yi2022add}.


Many works in ADD focus on finding proper audio features, which can be roughly categorized into hand-crafted and learning based features.
Hand-crafted features, although simple, have acceptable performance since their special design to extract audio properties.
For instance, the constant-Q transform (CQT) is good at capturing both long-range features and fine details in audio signals with its different filter window lengths across frequency bands\cite{das2019long}.
MFCC and LFCC features, have good match with human auditory characteristics and emphasis on low-frequency information to bolster speech detection tasks.
In recent years, the application of learning-based audio features in ADD tasks has attracted tremendous attention. Research has explored the use of audio features of Whisper \cite{radford2023robust} for detecting synthetic speech. Large amounts of audio data support the Whisper ASR system, demonstrating their superiority over handcrafted features \cite{kawa2023improved}. 
Similarly, self-supervised learning-based audio features have also proven beneficial for ADD tasks \cite{wang2021investigating}
The success of self-supervised models in various scenarios can be attributed to the usage of extensive pre-training data sourced from diverse domains, ensuring the model to produce meaningful audio features even in complicated situations.
These features aid in distinguishing between real and fake speech and perform well on out-of-domain dataset \cite{xie23c_interspeech}.

However, performance of ADD model based on single feature might be degraded since the spurious speech can be generated from distinctive audio synthesis systems, where single feature fails to represent characteristics of all the synthesis systems. 
Based on such phenomenon, we propose to use multiple features, which can improve the model generalizability by providing information from different aspects. Two methods are proposed based on feature selection and feature fusion respectively. These approaches can better capture the subtle differences between fake speech and real speech, bolstering detection system accuracy of identifying deep forgery samples especially generated by unknown synthesis systems.

This work focuses on improving the generalizability of ADD system, and contributions are:
\begin{enumerate}
    \item We investigate a broad range of handcrafted features and learning-based deep features. Experimental results show strong generalizability for 
    learning-based features pretrained on large amounts of data.
    \item We propose two multi-view feature incorporation methods to capture subtlety of the multiple candidate features to further improve the performance and generalizability of the system.
\end{enumerate}

\section{audio features and multi-view feature incorporation}
\label{sec:format}
In this section, we first introduce audio features of the hand-crafted and learning-based approaches, which are investigated in the experiments section.
Then, the proposed feature multi-view approaches based on feature selection and feature fusion are demonstrated.

\subsection{Audio features}
\subsubsection{Hand-crafted features}
Hand-crafted acoustic features have been well investigated in ADD studies.
In this paper, we evaluate 5 hand-crafted features including Mel-scaled spectrogram (Mel), Mel frequency cepstral coefficient (MFCC), log-frequency spectrogram (LogSpec), linear frequency cepstral coefficients (LFCC) and constant-Q transform (CQT).

\subsubsection{Learning-based features}
Learning-based acoustic features are generated from various audio tasks, and there is already precedent for using them for ADD tasks \cite{tak2021end,kawa2023improved,wang2021investigating}.
In this paper, 9 learning-based audio features proposed for various tasks have been extensively investigated and benchmarked for generalization performance on the ADD task.

The learnable acoustic front-ends automatically get the proper filter banks while optimizing the objective. We adopt SincNet \cite{ravanelli2018speaker} and LEAF \cite{zeghidour2021leaf} as learnable front-ends for ADD.
Besides, we also evaluate a range of deep learning-based audio features, where the use of additional data as well as task-related training approaches can be beneficial for the ADD task. 7 deep learning-based models across various tasks are chosen to generate audio features. For audio neural codec models, we use EnCodec \cite{defossez2022high} and AudioDec \cite{wu2023AudioDec} mainly consisted of autoencoder architecture and aimed to encode audio compactly. AudioMAE \cite{huang2022masked} is selected as representative of pretrained model towards universal audio perception. For pretrained model on human speech, we select Wav2Vec2, Hubert \cite{hsu2021Hubert}, and WavLM \cite{chen2022WavLM}, which share similar network architectures but different self-supervised losses. For the ASR model, we use Whisper \cite{radford2023robust} model trained on a large dataset from diverse speech scenario.

\subsection{Multi-view feature incorporation}
Features extracted from different deep models contain unique information, which can further boost ADD model generalizability with proper feature incorporation methods. Therefore, we propose two methods based on feature selection and feature fusion respectively.  

\subsubsection{Feature selection}
Identifying the most effective feature for the ADD task is difficult, especially for test data with unknown distributions. So, we introduce multi-feature candidates to improve the generalization of the ADD system. However, the introduction of redundant or irrelevant features 
may hinder the learning process of the classifier. Therefore, we propose a feature selection mechanism that decides whether to introduce a feature 
into the decision process based on sample-specific information, thus exploiting the information provided by multi-features while avoiding the negative impact of certain features.
\begin{equation}
    \label{eq:1}
    \begin{aligned}
        &\rm{m_{i}}= \mathcal{S}_{\bf{\theta}}(\mathrm{f_{i}})\\\
            &\rm{F_{select}}=\it{Concat}(\left\{\mathrm{f_i} \odot \rm{m_{i}}\right\}), i \in \left[0, N\right]
    \end{aligned}
\end{equation}
Our proposed feature selection mechanism is shown as eq\ref{eq:1}. Where $\mathrm{f_i}$ denotes the candidate features, $\rm{m_{i}}$ and $\rm{F_{select}}$ is the selected mask and select features respectively. Each feature goes through a selection module $\mathcal{S_{\theta}}$ before concatenated and fed into the classifier. This module consists of lightweight self-attentive layers, and the output of the module is a binary mask that determines whether the feature should be used in the decision for this sample. The discrete decision is obtained by the Gumbel-max method, thus allowing the selection module to be trained end-to-end with the whole system.

\subsubsection{Feature fusion}
Feature fusion, on the other hand, can incorporate all information in the multi-view feature without deleting any views.
To smoothly incorporate acoustic representations from different pretrained models, we combine channel attention mechanism and Transformer encoder to build a feature fusion module in (\ref{eq:2}). The multi-view feature, formed by concatenating candidate $\mathrm{f_i}$ on channel dimension, is first processed by a lightweight channel attention block to fuse on channel level (each channel represents one deep feature). 
Then, a Transformer encoder is applied to fuse the feature $\mathrm{r_i}$, on both time and frequency dimensions. With element-wise global receptive field, the final fused representation $\rm{F_{fusion}}$ are input into the classifier.
\begin{equation}
    \label{eq:2}
    \begin{aligned}
        &\rm{r_{i}}= {CA}(\it{Concat}(\mathrm{f_{i}}))\\\
            &\rm{F_{fusion}}={TE}(\it{Concat}(\left\{\mathrm{r_i}\right\})), i \in \left[0, N\right]
    \end{aligned}
\end{equation}
Where CA means channel-attention and TE means vallina Transformer encoder.


\section{Experiments}
\label{sec:pagestyle}

\subsection{Datasets}

We train our models on the train and dev subsets of the ASVSpoof 2019 Logical Access (LA) dataset part \cite{wang2020asvspoof}, which is consistent with most related works. To evaluate our systems, we adopt three datasets. The eval subset of ASVspoof 2019 and 2021 challenge are used to test the performance within similar domains \cite{yamagishi2021asvspoof}. 
The spoof audio of the ASVspoof challenge is generated by 11 TTS and 8 VC algorithms from VCTK corpus.
The samples of its eval subset is generated with different algorithms compared to the train subset. 
To evaluate the generalization ability of our systems, we also test our systems on In-the-Wild dataset, 
which contains 20.8 hours of real audio and 17.2 hours of deepfake audio\cite{muller2022does}. 
The In-the-Wild dataset is collected from the Internet and consists of audio from various realistic scenarios. 

\subsection{Implement details}

All audio samples are trimmed or padded to 4s and resampled to 16kHz for all acoustic features except the neural audio codec models EnCodec and AudioDec, which support sample rate of 24kHz.
For all handcrafted features, the window length and hop length are set to 25 ms and 10 ms, respectively. 

For speech self-supervised models, we employ the Wav2-Vec2 XLS-R \cite{babu2021xls} model pretrained on 128 languages,
Hubert-base model pretrained on LibriSpeech,
WavLM-Base-Plus model pretrained on Libri-Light,
GigaSpeech
and VoxPopuli
datasets. For the neural audio codec models, the continuous features of encoder output, instead of the discrete code, are used as audio features to prevent information loss. The AudioMAE model used in our experiments is pretrained on AudioSet.
The selected Whisper model is a tiny version pretrained on the speech recognition task. Besides, We select the 24khz version of the EnCodec and AudioDec model. Audio features extracted from above deep models are output of their encoders respectively. 

For all experiments, we use a ResNet18 as classifier. 
We train all of our systems with a cross-entropy loss. 
We use Adam optimizer with fixed learning rate at 1e-4 and weight decay at 1e-4. We train all the systems 100 epochs.
Checkpoint with lowest loss on validation set is saved for evaluation. All systems are evaluated by equal-error rate (EER).

\section{Result and analysis}
\label{sec:typestyle}

\subsection{Single feature}


\begin{table}[t]
    \centering
    \caption{Performance of various single audio features on the ADD task EER (\%)}
    \vspace{5pt}
	\label{tab:1}
	\begin{tabular}{@{}lccc@{}}
		\toprule
		Features   & \makecell{ASVspoof19 \\LA eval} & \makecell{ASVspoof21 \\DF eval} & In-the-Wild \\ \midrule
		Mel        & 7.42          & 20.13         & 50.56       \\
		MFCC       & 6.45          & 27.27         & 75.43       \\
		LogSpec    & 5.67          & 20.62         & 52.93       \\
		LFCC       & \hspace{-0.5em}15.35         & 25.67         & 65.45       \\
		CQT        & 4.91          & 20.75         & 56.69       \\
		LEAF       & 8.54          & 21.54         & 49.70       \\
		SincNet    & 6.12          & 20.78         & 56.74       \\
		EnCodec    & \hspace{-0.5em}10.25         & 24.93         & 39.44       \\
		AudioDec   & \hspace{-0.5em}10.47         & 26.13         & 43.69       \\
		AudioMAE   & 11.07          & 30.47         & 75.40       \\
		XLS-R      & \textbf{2.07}          & \textbf{11.78}         & 29.19       \\
		Hubert     & 6.78          & 14.76         & \textbf{27.48}       \\
		WavLM      & 7.24          & 15.53         & 30.50       \\
		Whisper    & 5.59          & 23.28         & 42.73       \\ \bottomrule
	\end{tabular}
	\vspace{-5pt}
\end{table}

Table \ref{tab:1} shows the results of our experiments, where we evaluate 14 audio features under the same experimental setup and test our systems on 3 datasets. The classification results for all features on the ASV2019 LA evaluation set are significantly better than those on the ASV2021 DF evaluation dataset and the In-the-Wild dataset. However, the results from the ASV2021 DF evaluation dataset and the In-the-Wild dataset show that the ADD system trained on the ASV2019 dataset is poorly generalized. 
The ASV2021 DF dataset contains samples from various spoofing systems that utilize different audio codec processing methods. For the In-the-Wild dataset, samples are collected from complex environments outside of professional studio and the speech content differs.

In our experiments, the handcrafted features fail to show reliable discrimination ability in realistic scenario. All the systems using handcrafted features get ERR greater than 50\% in the In-the-Wild dataset. The learnable front-end Leaf and SincNet learn filter banks during training, but still generalize poorly, with EERs 56.69 and 49.70 respectively.

On the contrary, most deep features show stronger generalizability.
The neural audio codecs EnCodec and AudioDec emphasize the compression rate and the fidelity of the decoded audio. While underperforming on the ASV2019 LA and ASV2021 DF evaluation datasets, these two models get an EER of 39.44 and 43.69 on the In-the-Wild dataset.
The Wav2Vec2 XLS-R model is pretrained on 436K hours of speech in 128 languages, based on which the system achieves the best EER on the ASV2019 LA and ASV2021 DF datasets. For results in the In-the-Wild dataset, EER decreases by 21.37 in comparison to the best manual features. 
The Hubert and WavLM features also perform excellently on the In-the-Wild dataset, where the Hubert feature achieves the best EER among all single-feature detection systems at 27.48.

Of all the deep features, the AudioMAE model pretrained on the audio spectrograms using mask autoencoder shows the poorest generalization on the ADD task. The EER is 75.40 on the In-the-Wild dataset, which is even worse than most of the handcrafted features. The failure might be attributed to the pretrained dataset. AudioMAE is pretrained on the Audioset dataset, which contains more universal audio than human speech, dispersing the ability to discriminate between true and false human speech.
The whisper feature also fails to generalize well, even though pretrained with more than 680k hours of unlabeled speech data. This feature is obtained by weakly supervised training on the ASR task, which focuses more on speech content instead of audio signal information.

\begin{figure}[tb]
	\label{fig:2}
	\begin{minipage}[b]{.48\linewidth}
		\centering
		\centerline{\includegraphics[width=4.0cm]{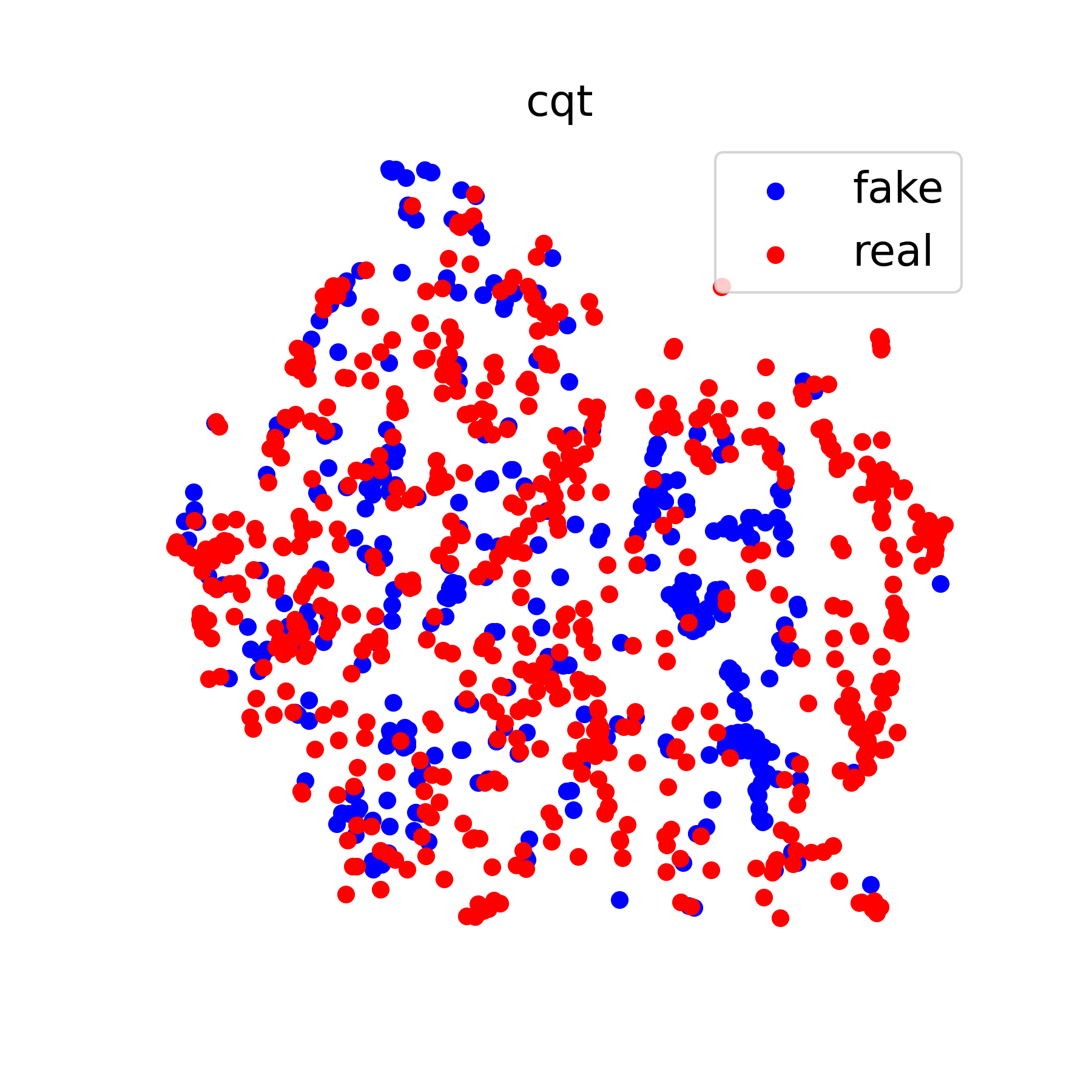}}
		\centerline{(a)}\medskip
	\end{minipage}
	\begin{minipage}[b]{0.48\linewidth}
		\centering
		\centerline{\includegraphics[width=4.0cm]{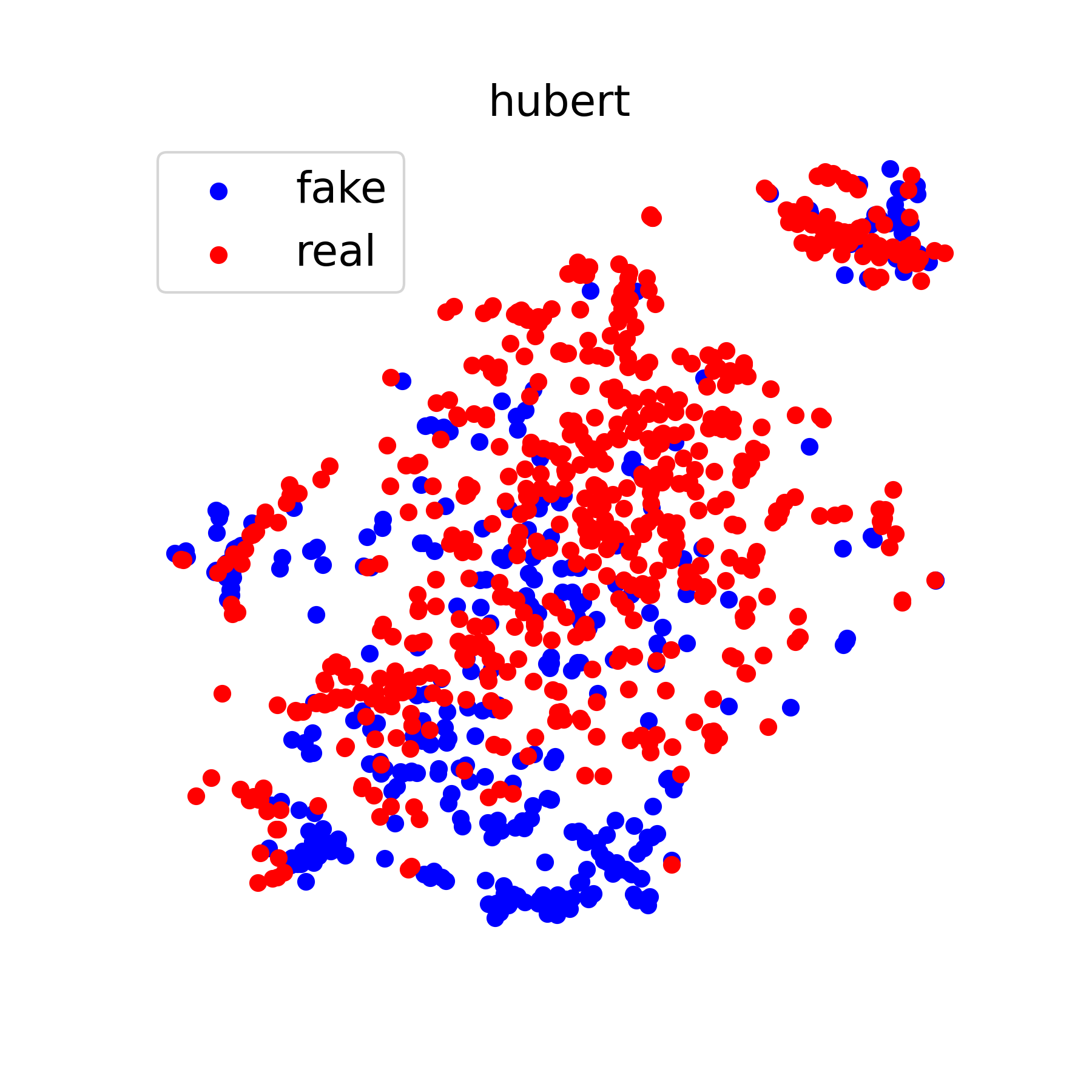}}
		\centerline{(b)}\medskip
	\end{minipage}
	\caption{Visualization of CQT features and Hubert features for real and fake speech in the In-the-Wild dataset.}
	\label{fig:res}
\end{figure}
To visualize the superior generalizability of deep features over hand-crafted features, Fig \ref{fig:2} shows the visualization of CQT features and Hubert features for real and fake speech in the In-the-Wild dataset using t-SNE. Although difficulty in discriminating between real and fake is revealed, the Hubert feature is more distinguishable than the CQT feature space where the two categories completely overlap.

\subsection{Multi-view feature incorporation}

Based on results from single-view feature experiments, Hubert, XLS-R, and WavLM features that perform well in the ASV2021 datasets are chosen as multi-view feature to further improve the generalizability of the detection system.


\begin{table}[t]
\centering
\caption{The comparison of EER (\%) score on In-the-Wild dataset.}
\vspace{5pt}
\label{tab:hard}
\begin{tabular}{@{}lcc@{}}
\toprule
\makecell[l]{Model}           &  \makecell[c]{Features}           & \makecell[c]{In-the-Wild}  \\ \midrule
RawNet2\cite{yi2023audio}         & waveform           & 36.74       \\
RawNet2\cite{muller2022does}         & waveform           & 33.94       \\
AASIST\cite{yi2023audio}          & waveform           & 34.81       \\
ResNet34\cite{yi2023audio}        & XLS-R              & 46.35       \\
LCNN\cite{yi2023audio}            & XLS-R              & 39.82       \\
Res2Net\cite{yi2023audio}         & XLS-R              & 36.62       \\ \midrule
ResNet18(ours)  & XLS-R              & 29.19       \\
ResNet18(ours)  & Hubert             & \textbf{27.48}       \\
ResNet18(ours)  & WavLM              & 30.50       \\ \midrule
Selection(ours) & XLS-R,WavLM,Hubert & 25.98            \\
Fusion(ours)    & XLS-R,WavLM,Hubert  & \textbf{24.27}            \\ \bottomrule
\end{tabular}
\vspace{-5pt}
\end{table}

Table \ref{tab:hard} shows results based on incorporating these three deep features on the In-the-Wild dataset. Compared to results either implemented in this work or results from studies \cite{yi2023audio,muller2022does}, both proposed approaches are proved beneficial to significantly improve the model generalizability where the EER reduces from 27.48, to 24.27 by the feature fusion, or to 25.98 by the feature selection. 
The effectiveness of feature selection comes from a sample-aware mask mechanism, based on which each individual sample can select the most appropriate feature, while single-feature detection system provide no feature selection space. The audio characteristic of the individual sample is learned to form the mask, which is also supervised by the detection task. This end-to-end approach guarantees such effectiveness.
On the other hand, the success of feature fusion indicates a complementary effect among the selected three deep features. Each value in the fused feature attends to any other value not only across time and frequency dimension, but also feature dimension. So, the fused feature are better representation for the ResNet18 classifier to get the best EER on the In-the-Wild dataset.

\section{Conclusion}
\label{sec:prior}
In this paper, we study the association between audio features and the generalizability of the ADD system. First, more audio features are tested and analyzed compared to any other studies on the ADD task, including handcrafted features, learnable audio front-end, audio neural codec, audio pretrained model, speech pretrained model, and speech recognition model in a total of 14 audio features. Experimental results show that in the In-the-Wild dataset, features of the speech pretraining models have good generalization performance while handcrafted features generalize poorly.
The generalization performance of speech features on ADD task comes from the large amount of pretraining data as well as the appropriate pretraining task. We further improve the generalization ability of the model based on the proposed feature selection and feature fusion methods. The results show that these two methods can improve the generalizability  compared to single features.

\section{Acknowledgement}
\label{sec:prior}
We gratefully acknowledge the support of MindSpore, CANN (Compute Architecture for Neural
Networks) and Ascend AI Processor used for this research.

\bibliographystyle{IEEEbib}
\bibliography{strings,refs}

\end{document}